\documentclass[12pt,notitlepage]{article}

\usepackage{amsmath}         
\usepackage{amssymb}         
\usepackage{amsfonts}        
\usepackage{graphicx}

\usepackage{color}         
\usepackage{slashed}       
\usepackage{framed}        
\usepackage{subcaption}    
\usepackage[mathscr]{euscript} 
\usepackage{cite}          
\usepackage{booktabs}      
\usepackage{youngtab}	    
\usepackage{arydshln} 	    
\usepackage{tocloft}		   
\usepackage{setspace}	   

\usepackage{listings}       
\usepackage[margin=2cm]{geometry}   
\graphicspath{{figures/}}	        
\numberwithin{equation}{section}    
\renewcommand{\tilde}{\widetilde}   

\newcommand{\email}[1]{\href{mailto:#1}{#1}}

\newenvironment{institutions}[1][2em]{\begin{list}{}{\setlength\leftmargin{#1}\setlength\rightmargin{#1}}\item[]}{\end{list}}

\let\oldenumerate\enumerate
\renewcommand{\enumerate}{
  \oldenumerate
  \setlength{\itemsep}{1pt}
  \setlength{\parskip}{0pt}
  \setlength{\parsep}{0pt}
}

\let\olditemize\itemize
\renewcommand{\itemize}{
  \olditemize
  \setlength{\itemsep}{1pt}
  \setlength{\parskip}{0pt}
  \setlength{\parsep}{0pt}
}

\usepackage{bbm}

\usepackage[
	colorlinks=true,
	citecolor=black,
	linkcolor=black,
	urlcolor=blue,
	hypertexnames=false]{hyperref}

\newcommand{\tev}{\ensuremath{\text{\small TeV}}}
\newcommand{\gev}{\ensuremath{\text{\small GeV}}}

\newcommand{\ZZ}{\ensuremath{\mathbbm{Z} }}

\newcommand{\barB}{\ensuremath{\overline{B}}}

\newcommand{\barM}{\ensuremath{\overline{M}}}
\newcommand{\barQ}{\ensuremath{\overline{Q}}}

\renewcommand{\P}{\barQ}

\newcommand{\tB}{\ensuremath{\mathcal B}}
\newcommand{\tM}{\ensuremath{\mathcal M}}
\newcommand{\tbarB}{\ensuremath{ \overline{\mathcal B} }}
\newcommand{\tbarM}{\ensuremath{ \overline{\mathcal M}} }

\newcommand{\abs}[1]{\ensuremath{\left|#1\right|}}

\newcommand{\yF}{{\ensuremath{\tiny\yng(1)}}}
\newcommand{\yFb}{{\ensuremath{\tiny \overline{\yng(1)}}}}

\newcommand{\rep}[1]{\ensuremath{{\bf #1}}}
\newcommand{\repbar}[1]{\ensuremath{\overline{\bf #1}}}

\newcommand{\mpl}{\ensuremath{M_\text{P}}}
\newcommand{\ev}[1]{\ensuremath{ \langle #1 \rangle} }
\renewcommand{\eqref}[1]{Eq.~(\ref{#1})}

\newcommand{\upq}{\ensuremath{U(1)_\text{PQ}}}

\newcommand\Tstrut{\rule{0pt}{2.5ex}}         
\newcommand\Bstrut{\rule[-1.3ex]{0pt}{0pt}}

\usepackage{xcolor}

\begin{document}

\begin{flushright}
UCI-HEP-TR-2018-12
\end{flushright}

\begin{center}

   {\Large \bf A High Quality Composite Axion}

    \vskip 1cm

    { \bf Benjamin~Lillard} and {\bf Tim~M.P.~Tait} 
    \\ \vspace{-.2em}
    { 
    \footnotesize
    \email{blillard@uci.edu},~
    \email{ttait@uci.edu}
    }
	
    \vspace{-.2cm}

    \begin{institutions}[2.25cm]
    \footnotesize
    {\it 
	    Department of Physics and Astronomy, University of California, Irvine, CA 92697, USA
	    }   
    \end{institutions}

\end{center}
\vspace*{0.5cm}

\begin{abstract}
 \noindent
 The strong CP problem is a compelling motivation for physics beyond the Standard Model.
The most popular solutions invoke a global \upq\ symmetry, but are challenged by quantum gravitational
corrections which are thought to be incompatible with global symmetries, arguing that realistic theories contain
additional structure.  We explore a construction in which the
 $\upq$ symmetry is protected to arbitrary order by virtue of a supersymmetric, 
 confining $SU(N)_L \times SU(N) \times SU(N)_R \times U(1)_X$ product gauge group, 
 achieving $\abs{\bar\theta} < 10^{-11}$ for an $SU(5)$ model with $f_a \lesssim 3 \times 10^{11}\, \gev$.
This construction leads to low energy predictions such as a 
$U(1)_X$ gauge symmetry, and for  $X = B-L$
engineers a naturally $\mathcal O(\tev)$ value for the $\mu$~parameter of the MSSM.
\end{abstract}


\section{Introduction}
\label{sec:intro}

Despite the conceptual simplicity of the axion solution to the strong CP problem, relatively few axion models have been developed which 
naturally predict $\abs{\bar\theta}\lesssim 10^{-11}$ when confronted with gravitationally induced \upq\ violating operators.
Models which do sufficiently protect the axion scalar potential from gravitational perturbations typically require large groups or complicated structures, leading to an ongoing search for more satisfying solutions.

In this work we present a relatively simple composite axion model in a confining supersymmetric theory, which is consistent with gauge coupling unification and compatible with current experimental results. Certain mesons in the theory are identified as composite Higgs fields, ameliorating the $B/\mu$ problem of the MSSM, and in one variant of our model the $B-L$ global symmetry of the Standard Model is gauged.

\subsection{The Strong CP Problem} \label{intro:CP}

The Standard Model (SM) contains several puzzles, one of the most pressing of which is the value of the $\theta$ parameter in the QCD Lagrangian:
\begin{equation}
\mathcal L = \frac{g^2\theta}{64 \pi^2 } \epsilon^{\mu\nu\rho\sigma} G_{\mu\nu}^a G_{\rho\sigma}^a  \equiv \frac{g^2}{32 \pi^2} \theta G_{\mu\nu} \tilde{G}^{\mu\nu}.
\end{equation}
Searches for an electric dipole moment of the neutron have so far resulted only in upper limits on its magnitude, implying
that $\abs{\bar\theta} < 6\times 10^{-11}$~\cite{Baker:2006ts,Afach:2015sja}, where $\bar\theta$ is the physically relevant combination of $CP$ violating phases,
\begin{equation}
\bar\theta = \theta + \text{arg\,}\text{det\,} M_Q,	 \label{eq:bartheta}
\end{equation}
where $M_Q$ is the quark mass matrix.
As the $\theta$ term violates both $P$ and $CP$, the unnaturally small value of $\bar\theta$ is referred to as the strong $CP$ problem.
For more complete reviews, see for example~\cite{Peccei:1977ur,Shifman:1979if,Kim:2008hd}.

In many popular solutions of the strong $CP$ problem, $\bar\theta$ is rendered unphysical by ensuring that the classical Lagrangian respects a global $U(1)$ symmetry, 
which is explicitly broken by the QCD anomaly. A simple example can be seen from \eqref{eq:bartheta}: if one sets $m_u = 0$: an axial $U(1)_A$ symmetry 
emerges in this limit, so that $\text{arg\,}\text{det\,} M_Q$ (and therefore $\bar\theta$) becomes unphysical. 
If it were not for compelling evidence that $m_{u,d} \neq 0$, this ``massless up quark solution" would 
naturally explain the absence of $CP$ violation in the strong sector.

Axion models address the strong $CP$ problem by associating $\bar\theta$ with the pseudo-Nambu--Goldstone boson of an approximate \upq\ global symmetry. 
This is achieved by introducing a (SM singlet)
complex scalar $\phi$ together with left-handed color (anti)-triplet fermions $Q$ and $\barQ$, along with the interaction
\begin{equation}
\mathcal L \supset V(\phi) + \phi Q \barQ + h.c. 
\end{equation} 
where $V(\phi)$ is designed such that $\phi$ acquires an expectation value $\ev{\phi} \gtrsim 10^9\, \gev$. 
The bare mass term $m Q \barQ$ is forbidden, so that $\mathcal L$ respects a \upq\ symmetry under which $\phi$ is charged. 
The $SU(3)_c^2$-$\upq$ anomaly coefficient is nonzero, as can be seen from the fact that $(Q \barQ)$ carries a net \upq\ charge.

Expanding about the $\ev{\phi}\neq0$ vacuum, the axion $a$ is identified as the phase of $\phi$:
\begin{equation}
\phi = \left( \ev{\phi} + \frac{\sigma}{\sqrt{2}} \right) \exp\left( i \frac{a}{f_a} \right),
\end{equation}
where $f_a \equiv \sqrt{2} \ev{\phi}$.
The $SU(3)_c^2$-\upq\ anomaly induces an $a G \tilde{G}$ coupling,
\begin{equation}
\mathcal L = \frac{g^2}{32\pi^2} \left( \bar\theta - \frac{a}{f_a} \right) G_{\mu\nu} \tilde{G}^{\mu\nu},
\end{equation}
and nonperturbative QCD dynamics generate a periodic potential for $a$ which can be heuristically 
(up to chiral symmetry-violating corrections~\cite{DiVecchia:1980yfw}, which are unimportant for our discussion)
described by
\begin{equation}
V(a) \simeq m_\pi^2 f_\pi^2 \left(1 - \cos\left[ \frac{a}{f_a} - \bar\theta \right] \right). \label{eq:Va}
\end{equation}
The axion potential is minimized by $\ev{a} = f_a \bar\theta$, so that $CP$ is conserved in the QCD vacuum. 

In ``invisible axion" models of this type~\cite{Kim:1979if,Shifman:1979if} the axion is light and weakly coupled, with a mass given by:
\begin{equation}
m_a^2 \simeq \frac{m_\pi^2 f_\pi^2}{f_a^2}.	\label{eq:axionmass}
\end{equation}
A lower bound $f_a \gtrsim 10^9\,\gev$ is set primarily by astrophysical observations of stellar cooling and supernovae.
In much of the parameter space, the axion provides a natural dark matter candidate:
its interactions are suppressed by the decay constant $f_a$, and it can be produced in the early universe by the misalignment mechanism~\cite{Preskill:1982cy}.
For $\mathcal O (1)$ initial misalignment angles, the correct relic abundance is obtained for $f_a \lesssim 10^{12}\,\gev$, though $f_a$ could
be larger if the misalignment was smaller.
The fact that the QCD axion could also play the role of dark matter is one of the reasons for its continued popularity as a solution to the strong $CP$ problem. 

\subsection{Axion Quality Problem}

A closer inspection of the simple axion model presented above reveals a new set of theoretical difficulties, namely a hierarchy problem and a fine-tuning problem. 
The axion model prefers on a hierarchy between the scale of symmetry breaking $f_a$ and the Planck mass, $\mpl$. 
A number of standard solutions, such as supersymmetry or compositeness, have been proposed which would 
render an axion scale $f_a \ll \mpl$ technically natural. 
However, many axion models still suffer from a more severe fine-tuning, known as the \emph{axion quality problem}.

Arguments from general relativity~\cite{Giddings:1987cg,Kamionkowski:1992mf,Barr:1992qq,Kallosh:1995hi,Abbott:1989jw,Coleman:1989zu} 
suggest that non-perturbative quantum gravitational effects do not respect global symmetries such as baryon number or \upq. 
This is highly problematic for most axion models, which rely 
on \upq\ being an exact symmetry in the $\alpha_s \rightarrow 0$ limit, explicitly broken only by the QCD anomaly. 
If additional PQ-violating operators representing the short distance influence of quantum gravity such as
\begin{equation}
\Delta V(\phi) =  \frac{\abs{\phi}^{k+3} }{\mpl^{k}} \left(\lambda_k \phi +\lambda_k^\star \phi^\star\right)
\end{equation}
are present, the corresponding perturbation in $V(a)$ can shift $\ev{a}$ far away from the $CP$-conserving value of \eqref{eq:Va}:
\begin{equation}
\delta V(a) \sim \lambda_k f_a^4 \left(\frac{f_a}{\mpl}\right)^k \cos\left(\Delta_\text{PQ} \frac{a}{f_a} - \varphi \right), \label{eq:Vap}
\end{equation}
where the phase $\varphi$ is determined by $\lambda_k$, and $\Delta_\text{PQ}$ is the \upq\ charge of the operator $\phi$. 
It is convenient to describe such perturbations by defining a ``quality factor" $Q$:
\begin{equation}
\delta V(a) = Q f_a^4 \cos\left(\frac{a}{f_a} - \varphi \right).
\end{equation}
If we assume $\varphi \sim \mathcal O(1)$ is not tuned, the measured value of $\abs{\bar\theta} \lesssim 10^{-11}$ is possible only if $\delta V(a)$ satisfies
\begin{equation}
Q \lesssim 10^{-63} \left(\frac{10^{12}\, \gev}{f_a} \right)^4. \label{eq:quality}
\end{equation}
Satisfying this bound requires that the theory of quantum gravity somehow produce a severe fine-tuning in the $\lambda_k$, 
such that even the dimension-12 operators in \eqref{eq:Vap} must have $\lambda_k \ll 1$.

In a truly compelling axion model, the \upq\ symmetry should emerge 
as a consequence of some other underlying structure which forbids the problematic operators. 
For example, a gauged discrete $\ZZ_{n}$ symmetry~\cite{Chun:1992bn} for some $n \gtrsim 13$ can forbid all PQ-violating operators smaller than $(\phi^n + c.c.)$. 
Composite axion models such as~\cite{Randall:1992ut,DiLuzio:2017tjx,Lillard:2017cwx} also protect \upq\ to arbitrarily high order, 
with the added benefit that the axion scale $f_a$ can be generated dynamically.
Other constructions~\cite{Cheng:2001ys,Fukuda:2017ylt} associate \upq\ with a different, gauged $U(1)$, so that many of the PQ-violating operators are forbidden. 
Many of these constructions are intricate and also rather delicate in the sense that the axion quality is easily ruined in extensions of the model.

In this work we present an alternative composite axion model based on an $SU(N)\times SU(N)$ confining supersymmetric gauge theory with simple matter content. 
The Standard Model matter fields and interactions are easily embedded, and we show that the axion quality is preserved even with the addition of new fields. Upon identifying the $H_u$ and $H_d$ doublets as mesons from $SU(N)$ confinement, we find that the $\mu$ parameter of the MSSM naturally assumes an $\mathcal O(\tev)$ value. Finally, we explore the ability of this model to mediate supersymmetry breaking via composite messengers.

\section{Composite Axion Model} 
\label{sec:axionmodel}

Conjectured dualities~\cite{Seiberg:1994bz,Intriligator:1994sm} allow one to analyze the low energy behavior of supersymmetric gauge theories. 
In particular, an $SU(N_c)$ gauge theory with $N_f = N_c$ flavors of quarks $(Q+\barQ)$ in the (anti-)fundamental representation is expected to confine at a 
characteristic scale $\Lambda$, such that the low energy degrees of freedom are described by the gauge-singlet operators
\begin{align}
M = (Q \barQ), &&
B = (Q^N), &&
\barB = (\barQ^N),
\end{align}
subject to the quantum-modified constraint
\begin{equation}
\det M - B \barB = \Lambda^{2N}. \label{eq:qdms}
\end{equation}
The constraint \eqref{eq:qdms} guarantees that the global $SU(N_f) \times SU(N_f) \times U(1)$ symmetry is spontaneously broken, either by $\ev{M}\neq 0$ or $\ev{B\barB}\neq 0$.
Similar behavior has been demonstrated in theories with product gauge groups of the form $SU(N)\times SU(N)\times \ldots \times SU(N)$ with bifundamental matter~\cite{Chang:2002qt}. We show that a composite axion emerges in a subset of these theories, with sufficiently high axion quality.

\begin{figure}
\includegraphics[scale=0.8]{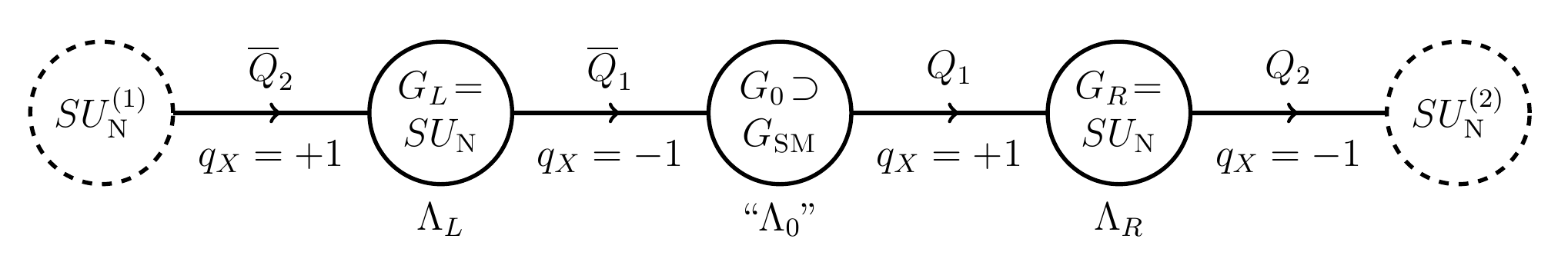}
\caption{Moose diagram indicating the charges of bifundamental matter fields $\barQ_{1,2}$ and $Q_{1,2}$ under the gauge group 
$SU(N)_L \times SU(N)_{SM} \times SU(N)_R \times U(1)_X$ and global $SU(N)_1 \times SU(N)_2$ global symmetries. 
The Standard Model $SU(3)_c\times SU(2)_L \times U(1)_Y$ is a subgroup of $G_0$.}
\label{fig:moose}
\end{figure}

We invoke the gauge group $SU(N)_L \times SU(N)_{SM} \times SU(N)_R \times U(1)_X$, where $SU(N)_{SM}$ contains the Standard Model $SU(3)_c \times SU(2)_L \times U(1)_Y$ either as a gauged subgroup
or as an $SU(5)$ grand unified theory. 
The strongly coupled $SU(N)_{L,R}$ confine at the characteristic scales $\Lambda_{L,R} \gg \tev$, 
but the Abelian $U(1)_X$ is weakly coupled\footnote{The axion construction leaves the charges of the MSSM matter under
$U(1)_X$ largely undetermined.  We explore several alternatives below.}. 
The bifundamental fields $\barQ_{1,2}$ and $Q_{1,2}$ have $U(1)_X$ charges $\pm 1$, 
as depicted in the moose diagram of Figure~\ref{fig:moose}, with \upq\ charges shown in Table~\ref{table:UVtheory}.

Below the scales $\Lambda_L$ and $\Lambda_R$, the low energy degrees of freedom are described by the composite operators
satisfying  equations of motion:
\begin{align}
\def\spc{0.5cm}
\begin{array}{c c c c c c l}
\barM = (\P_2 \P_1) &\hspace{\spc}&
\barB_1 = (\P_1^N) &\hspace{\spc}&
\barB_2 = (\P_2^N) &\hspace{\spc}&
\Lambda_L^{2N} = \det \barM - \barB_1 \barB_2   \\ \vspace{-0.2cm} \\
M = (Q_1 Q_2) &&
B_1 = (Q_1^N) &&
B_2 = (Q_2^N) &&
\Lambda_R^{2N} = \det M - B_1 B_2.
\end{array}
\label{eq:IRdof}
\end{align}
In the absence of a superpotential, this model respects the global $SU(N)_1\times SU(N)_2$ symmetries shown in Figure~\ref{fig:moose}, as well the gauged $U(1)_X$.
There is also a conserved $U(1)_R$, under which the gauginos have charge $+1$ and all of the $\barQ_{1,2}$ and $Q_{1,2}$ are neutral, which remains unbroken 
everywhere on the moduli space.

In the regime where $G_0$ is weakly coupled, there is another nearly exact global symmetry, \upq, which is broken only by the $G_0^2$-\upq\ anomaly. 
Due to the locally conserved $U(1)_X$, there is no unique assignment of Peccei--Quinn charges: rotations 
under \upq\ can always be combined with a global $U(1)_X$ transformation to define a new, equally valid Peccei--Quinn symmetry. 
This degeneracy is parameterized by the parameter $\alpha$ in Table~\ref{table:UVtheory}.

\begin{table}
\centering
\hspace*{-0.75cm}
\begin{tabular}{| c | c | c c c | c | c | c |} \hline
\Tstrut \Bstrut
   	&$SU(N)_1$&$G_L$&$G_0$	&$G_R$&$SU(N)_2$	&$U(1)_X$& $\upq$	\\ \hline \Tstrut
$\P_2$  &\yF	&	\yFb	&		&		&			&	$1$	& 	$-(1-\alpha)/N$	\\ 
$\P_1$&		&	\yF	&	\yFb	&		&			&	$-1$	& 	$(1-\alpha)/N$		\\
$Q_1$&		& 		&	\yF	&	\yFb	&			&	$1$	&	$(1+\alpha)/N$		\\ 
$Q_2$&		& 		&		&	\yF	&	\yFb		&	$-1$	&	$-(1+\alpha)/N$		\\ \hline \Tstrut
$\barM$&	\yF	&		&	\yFb	&		&			&	0		&	0		\\ 
$M$	&		&		&	\yF	&		&	\yFb		&	0		&	0		\\
$\barB_2$	&	&		&		&		&			&	$N$		&	$-1+\alpha$		\\
$\barB_1$	&	&		&		&		&			&	$-N$		&	$1-\alpha$		\\
$B_1$	&	&		&		&		&			&	$N$		&	$1+\alpha$		\\
$B_2$	&	&		&		&		&			&	$-N$		&	$-1-\alpha$		\\ \hline
\end{tabular}
\caption{\upq\ charges and representations under the gauged $G_L \times G_0 \times G_R$ and the global $SU(N)_1 \times SU(N)_2$ symmetries
are indicated
for the bifundamental quarks (upper half) and composite operators resulting from $G_L \times G_R$ confinement (lower half).
}  
\label{table:UVtheory}
\end{table}

On the quantum-deformed moduli space described by \eqref{eq:IRdof}, the global $SU(N)_1 \times SU(N)_2 \times U(1)_X \times \upq$ symmetry must be 
broken to a subgroup. Furthermore, if the low energy limit of this theory is to approach the Standard Model, then it must be true that $\det M = \det \barM = 0$; 
otherwise, $SU(3)_c$ would be broken in the vacuum.
The vacuum therefore must be engineered to lie on the $\ev{B_1 B_2} \neq 0$, $\ev{\barB_1 \barB_2} \neq 0$ branch of the moduli space, 
where $U(1)_X$ and \upq\ are both spontaneously broken, and the $U(1)_X$ vector supermultiplet acquires a mass by ``eating" a 
combination of the chiral superfields.  This is accomplished by including a term in the superpotential of the form:
\begin{eqnarray}
\frac{\left( \overline{Q}_2 \overline{Q}_1 \right) \left( Q_1 Q_2 \right)}{M_*}
\end{eqnarray}
which after confinement generates a mass term for the mesons, 
$W \sim \mu \barM M$, lifting the mesonic flat directions.  If not otherwise present, this term is expected to be induced
by quantum gravitational effects.

A unique definition of the Peccei--Quinn charges emerges once $U(1)_X$ is broken: by canonically normalizing the 
kinetic terms of the (would-be) Nambu--Goldstone bosons of \upq\ and $U(1)_X$, the parameter $\alpha$ of Table~\ref{table:UVtheory} is
related to the vacuum expectation values (VEVs) of the baryons as
\begin{equation}
\alpha = \frac{\bar v_1^2 + \bar v_2^2 - v_1^2 - v_2^2}{f_X^2},
\end{equation}
where
\begin{align}
\bar v_i^2 = 2 \abs{ \frac{\ev{\barB_i}}{\Lambda_L^{N-1}} }^2 &,&
 v_i^2 = 2 \abs{ \frac{\ev{B_i}}{\Lambda_R^{N-1}} }^2 &,&
f_X^2 = \bar v_1^2 + \bar v_2^2 + v_1^2 + v_2^2,
\label{eq:fXetc}
\end{align}
and where the axion decay constant $f_a$ is 
\begin{equation}
f_a^2 = f_X^2 \left( 1 - \alpha^2 \right).
\end{equation}
With this normalization, a \upq\ rotation by a phase $\theta$ is achieved by the linear shift
\begin{equation} 
a \rightarrow a + \theta f_a. \label{eq:ashift}
\end{equation}
Although the products $v_1 v_2$ and $\bar v_1 \bar v_2$ are set by the quantum modified constraints,
\begin{align}
\bar v_1 \bar v_2 = 2 \abs{ \Lambda_L^2 }&,&
 v_1  v_2 = 2 \abs{ \Lambda_R^2 },
 \end{align}
the values of the decay constants $f_a$ and $f_X$ vary along the flat directions within the allowed ranges
\begin{align}
f_X^2 \geq 4  \abs{\Lambda_L^2} + 4  \abs{\Lambda_R^2} &,&
f_a^2 \leq f_X^2.
\end{align}
The case $f_a \ll f_X$ is achieved in the limits $\Lambda_L \gg \Lambda_R$ or $\Lambda_L \ll \Lambda_R$, 
as $\alpha \rightarrow \pm1$. Conversely, the special case $v_1^2 + v_2^2 = \bar v_1^2 + \bar v_2^2$ corresponds to $f_a = f_X$.

\subsection{Axion Quality}

To examine the axion quality, we introduce operators characterized by $\mpl$ which represent an
effective field theory description of the low energy residual effects of
quantum gravity.
It is convenient to introduce a set of rescaled composite operators with mass dimension $+1$:
\begin{align}
\tbarM = \frac{(\barQ_2 \barQ_1)}{\Lambda_L} 
&& \tM = \frac{(Q_1 Q_2)}{\Lambda_R} 
&& \tbarB_i = \frac{(\barQ_i^N)}{\Lambda_L^{N-1}} 
&& \tB_i = \frac{(Q_i^N)}{\Lambda_R^{N-1}}.
\end{align}
The effective gravitational superpotential violating all of the global symmetries takes the form:
\begin{align}
W_g &= \lambda_1 \frac{(\barQ_1^N)(Q_1^N)}{\mpl^{2N-3}} + \lambda_2 \frac{(\barQ_2^N)(Q_2^N)}{\mpl^{2N-3}} + \lambda_3 \frac{(\barQ_2^N)(\barQ_1^N)}{\mpl^{2N-3}} + \lambda_4 \frac{(Q_1^N)(Q_2^N)}{\mpl^{2N-3}} 
+ \rho_1 \frac{(\barQ_2 \barQ_1) (Q_1 Q_2)}{\mpl}
+ \ldots \\
&= \left( \frac{\Lambda_L^{N-1} \Lambda_R^{N-1}}{\mpl^{2N-3}} \right)\left\{ \lambda_1 \tbarB_1 \tB_1 + 
 \lambda_2 \tbarB_2 \tB_2 + \lambda_3 \tbarB_1 \tbarB_2 + \lambda_4 \tB_1 \tB_2 \right\} 
 + \rho_1 \left(\frac{\Lambda_L \Lambda_R}{\mpl} \right) \tbarM \tM + \ldots,
\label{eq:wgPQ}
\end{align}
with parameters $\lambda_i$ and $\rho_i$ encoding the UV physics.
Of the operators listed above, only the two associated with $\lambda_1$ and $\lambda_2$ violate \upq. All of the lower-dimensional operators such as $(\barQ_2 \barQ_1) (Q_1 Q_2)$ are neutral under \upq, and thus not harmful to the axion quality.

In a supersymmetric vacuum, the leading \upq\ violation appears with $\mpl^{4N-6}$ suppression in the Lagrangian: for example, within terms such as
\begin{equation}
\abs{ \frac{\partial W_g}{\partial \tB_1} }^2 = \abs{\frac{ \Lambda_L^{N-1} \Lambda_R^{N-1}}{\mpl^{2N-3}}}^2 \abs{  \lambda_1 \tbarB_1 + \lambda_4 \tB_2 }^2,
\end{equation}
implying a perturbation to the axion potential on the order of
\begin{equation}
Q f_a^4 \sim  \abs{\lambda_1 \lambda_4} \left( \frac{\sqrt{\Lambda_L \Lambda_R}}{\mpl} \right)^{4N-4} \mpl^2 \ev{\tbarB_1}\ev{\tB_2}.
\end{equation}
Taking $\Lambda_{L} \approx \Lambda_R \approx f_a \approx 10^{11}\, \gev$ as a benchmark and ignoring $\mathcal O(1)$ factors, the quality factor
\begin{equation}
Q \sim \abs{ \lambda_1 \lambda_4} 10^{48-32 N}
\end{equation}
satisfies the bound given in \eqref{eq:quality} for $N> 3$, even when the $\lambda_i$ are $\mathcal O(1)$.

More serious perturbations to the axion potential emerge when supersymmetry breaking is taken into account. Supersymmetry breaking induces an ``$A$-term" potential,
\begin{equation}
-\mathcal L_A = \left( \frac{\Lambda_L^{N-1} \Lambda_R^{N-1}}{\mpl^{2N-3}} \right)\left\{ A_1 \lambda_1 \tbarB_1 \tB_1 + 
 A_2 \lambda_2 \tbarB_2 \tB_2 + A_3 \lambda_3 \tbarB_1 \tbarB_2 + A_4 \lambda_4 \tB_1 \tB_2 \right\} + h.c, 
\label{eq:APQV}
\end{equation}
where the mass scales $A_i$ are in principle calculable once a particular mechanism of supersymmetry breaking is specified.
To remain agnostic concerning the details of supersymmetry-breaking,
we assume that the $A_i$ should be of roughly the same magnitude as the $SU(3)_c\times SU(2)_L \times U(1)_Y$ gaugino masses.

Both the $A_1$ and $A_2$ terms in \eqref{eq:APQV} perturb the axion potential:
\begin{equation}
\delta V(a) = 2  \frac{\Lambda_L^{N-1} \Lambda_R^{N-1}}{\mpl^{2N-3}} \left\{ \abs{ A_1 \lambda_1 \ev{\tbarB_1} \ev{\tB_1} }  \cos \left( 2\frac{ a}{f_a} + \varphi_1 \right) + \abs{ A_2 \lambda_2 \ev{\tbarB_2} \ev{\tB_2} }  \cos \left( 2\frac{ a}{f_a} + \varphi_2 \right) \right\}.
\end{equation}
Again taking $\Lambda_{L,R} \approx f_a \approx 10^{11}\, \gev$,
the constraint on the quality factor \eqref{eq:quality} can be written as
\begin{equation}
\frac{\lambda_i A_{i} }{10^4\, \gev} \left(\frac{10^{19}\, \gev}{\mpl}\right)^{2N-3} \left(\frac{\Lambda_L \Lambda_R}{10^{22}\, \gev^2} \right)^{N-1} \frac{\ev{\tbarB_{i}} \ev{\tB_{i}}}{10^{22}\, \gev^2} \cdot  10^{-16 N} \lesssim 10^{-76}
\label{eq:msQ}
\end{equation}
for $i=1,2$,
indicating that models with $N \geq 5$ are free from fine-tuning as long as the characteristic scales $\Lambda_{L,R}$ and $f_a$ are not much larger than $10^{11}\, \gev$.

\begin{figure}[h]
\centering
\includegraphics[width=0.94\textwidth]{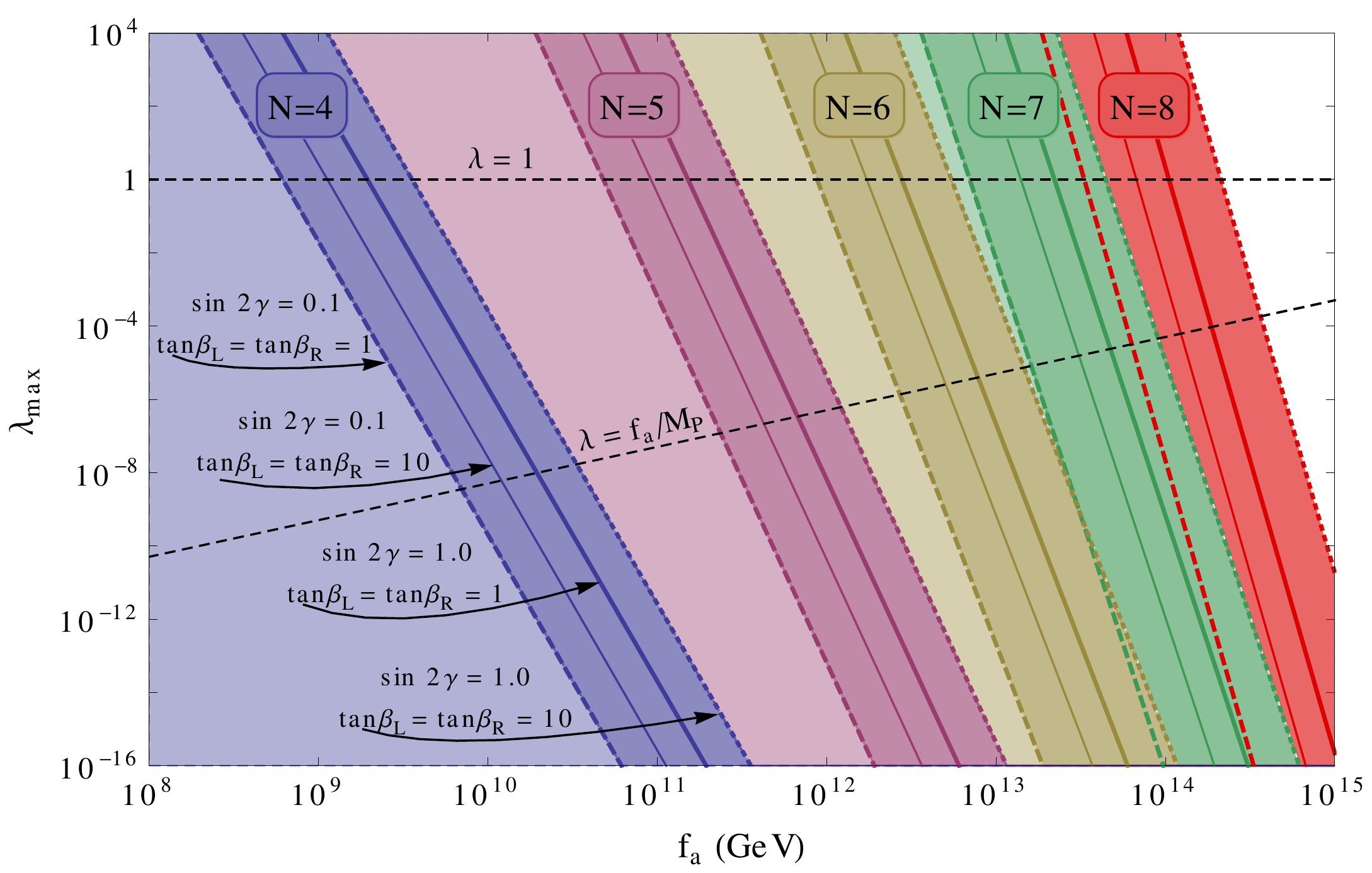}
\caption{
Maximum values of $\lambda_1 \approx \lambda_2$ consistent with \eqref{eq:newQ} for given values of $f_a$ and $N=4,5,6,7,8$. 
The region to the left of each line indicates the axion models which return $\abs{\bar\theta} < 10^{-11}$ without any fine tuning. 
From left to right within each band of a given $N$, models are indicated with: $\sin2\gamma=0.1$, $\tan\beta_L= \tan\beta_R =1$ (thin, dashed); 
$\sin2\gamma=0.1$, $\tan\beta_L=\tan\beta_R =10$ (thin, solid); $\sin2\gamma = \tan\beta_L = \tan\beta_L = 1$ (thick, solid); and 
$\sin2\gamma = 1$,  $\tan\beta_L = \tan\beta_L = 10$ (thin, dotted).
In each case $A_1 \approx A_2 = 10^5\, \gev$.
\label{fig:quality}
}
\end{figure}

In Figure~\ref{fig:quality} we plot the maximum values of $\lambda_i$ consistent with \eqref{eq:msQ}, for given values of $f_a$, $N$, and the other parameters, with the simplifying assumptions $A_1 \approx A_2$ and $\lambda_1 \approx \lambda_2$. It is convenient to label the vacua with the following parameterization:
\begin{align}
\tan \beta_L = \frac{\bar v_2}{\bar v_1} &&
\tan \beta_R = \frac{v_2}{v_1} &&
\sin^2 2 \gamma = \frac{f_a^2}{f_X^2} = 1 - \alpha^2.
\end{align}
All of the dimensionful parameters except for $A_i$ and $\mpl$ are now expressed in terms of $f_a$:
\begin{align}
\bar v_1 =  \frac{\cos \beta_L}{2 \cos \gamma} f_a && \bar v_2 =  \frac{\sin \beta_L}{2 \cos \gamma} f_a &&
v_1 =  \frac{\cos \beta_R}{2 \sin \gamma} f_a && v_2 =  \frac{\sin \beta_R}{2 \sin \gamma} f_a,
\end{align}
so that the axion quality condition is expressed:
\begin{equation}
\frac{Q f_a^4}{\mpl^4} = 8 \left(\frac{f_a^2}{8 \mpl^2 \sin 2 \gamma} \right)^N \left( \sin 2\beta_L  \sin 2 \beta_R  \right)^{\frac{N-1}{2}}  \left( \frac{\lambda_1 A_1 \cos \beta_L \cos \beta_R +\lambda_2 A_2 \sin \beta_L \sin \beta_R}{\mpl}  \right) \lesssim 10^{-88}.
\label{eq:newQ}
\end{equation}

Because $\beta_{L,R}$ label degenerate vacua on the moduli space defined by \eqref{eq:IRdof},
particularly large or small values of $\tan \beta_{L,R}$ are typically unnatural. 
On the other hand, $\gamma$ is primarily determined by the ratio $\Lambda_L/\Lambda_R$:
\begin{equation}
\tan \gamma = \frac{\Lambda_L}{\Lambda_R} \sqrt{ \frac{\sin 2 \beta_L}{\sin 2 \beta_R}  },
\end{equation}
so large or small values of $\tan \gamma$ are more easily tolerated from a naturalness perspective. 
As we see from \eqref{eq:newQ}, the best axion quality is achieved for $\tan \gamma \approx 1$, when $f_a \approx f_X$ and $\Lambda_L \approx \Lambda_R$.

We show the maximum tolerable $\lambda_1 \approx \lambda_2$ as a function of $f_a$ for a few choices of $N$, $\tan \beta_L = \tan \beta_R$,
and $\sin 2\gamma$ in Figure~\ref{fig:quality}.  
While effective field theory would suggest that generic theories of quantum gravity should produce $\lambda_{1,2} \sim \mathcal O(1)$,
in~\cite{Abbott:1989jw,Coleman:1989zu,Kallosh:1995hi} it is argued that wormhole-induced \upq\ violation yields suppressed values of 
$\lambda_i \sim \exp(-S_w)$, where the wormhole action $S_w$ depends logarithmically on the axion decay constant, $S_w \sim a - b \ln \frac{f_a}{\mpl}$. 
For typical cases the resulting suppression in $\lambda_i$ is modest: values as small as $\lambda\sim 10^{-7}$ 
are achieved in~\cite{Kallosh:1995hi} for $f_a \sim 10^{12}\, \gev$.
For $N=5$ such that $G_0$ is large enough to contain the SM, $\mathcal O(1)$ $\lambda$'s are consistent with $f_a \lesssim 10^{11}$~GeV.

Generally, the high axion quality observed in \eqref{eq:msQ} is preserved even when new fields are coupled to the model
provided that they are neutral under $U(1)_X$.  
Problems arise if there are fields $S$ with $U(1)_X$ charges:
\begin{equation}
q_S = \pm N, \pm \frac{N}{2}, \pm \frac{N}{3} , \ldots, \pm \frac{N}{N-1}, \label{eq:badQ}
\end{equation}
for which case $W_g$ includes gauge-invariant terms $S^p B_{1,2}$ or $S^p \barB_{1,2}$ for some power $p< N$. 

\subsection{$U(1)_{B-L}$ as $U(1)_X$} 
\label{sec:BL}

From \eqref{eq:msQ} we see the remarkable fact that for $f_a \lesssim 10^{11}\, \gev$ and
$\mathcal O(1)$ values in the couplings $\lambda_i$,  sufficient protection of the axion quality 
requires $N \geq 5$: precisely the right size to fit the entire Standard Model within $G_0$.
In this section we take $G_0 = SU(5)$ to be a global symmetry with a gauged $SU(3)_c \times SU(2)_L \times U(1)_Y$ subgroup, and we identify $U(1)_X$ as the $B-L$ symmetry of the Standard Model. 
The mesons $\tM (\rep{5})$ and $\tbarM(\repbar{5})$ decompose into irreducible representations of 
$SU(3) \times SU(2)\times U(1)$:
\begin{align}
\tM(\rep{5}) & \longrightarrow \tM^{(3)} (\rep{3},\rep{1})_{-\frac{1}{3}} \oplus \tM^{(2)} (\rep{1},\rep{2})_{\frac{1}{2}} \\
\tbarM(\repbar{5}) & \longrightarrow \tbarM^{(3)} (\repbar{3},\rep{1})_{\frac{1}{3}} \oplus \tbarM^{(2)} (\rep{1},\rep{2})_{-\frac{1}{2}} .
\end{align}
Table~\ref{table:BLuv} indicates the representations of the composites 
under the SM, plus three generations of MSSM matter and three right-handed neutrinos 
necessary to cancel the $U(1)_{B-L}$ gauge anomaly.

\begin{table}[t]
\centering
\hspace*{-0.75cm}
\begin{tabular}{| c | c | c c c | c | c | c |} \hline
\Tstrut \Bstrut
   		&$SU(5)_1$	&$SU_3$&$SU_2$&$U(1)_Y$	&$SU(5)_2$	&$U(1)_{B-L}$	& \upq			\\ \hline \Tstrut
$\tbarM^{(3)}$&	\rep{5}	&\repbar{3}&		& $1/3$	&			&$0$			&	0		\\ 
$\tbarM^{(2)}$&	\rep{5}	&		&\rep{2}	& $-1/2$	&			&$0$			&	0		\\ \hline \Tstrut
$\tM^{(3)}$&			&\rep{3}	&		& $-1/3$	&	\repbar{5}	&	$0$		&	0		\\ 
$\tM^{(2)}$&			&		&\rep{2}	& $1/2$	&	\repbar{5}	&	$0$		&	0		\\ \hline
$Q_L$	&			& \rep{3}	&\rep{2}	&$1/6$	&			& $+1/3$		&	0			\\
$\bar u_R$&			& \repbar{3}&		&$-2/3$	&			& $-1/3$		&	0			\\
$\bar d_R$&			& \repbar{3}&		&$1/3$	&			& $-1/3$		&	0			\\ \hline
$L$		&			& 		&\rep{2}	&$-1/2$	&			&	$+1$		&	0			\\
$\bar e_R$&			&		&		&$+1$	&			&	$-1$		&	0			\\
$\bar \nu_R$&			&		&		& $0$	&			&	$-1$		&	0		\\ \hline \Tstrut
 $\tB_1$, $\tbarB_2$ &	&		&		&	0	&			&	$5 q$	&$\pm 1 + \alpha$ 		\\
$\tbarB_1$, $\tB_2$ &	&		&		&	0	&			&	$-5 q$	&$\pm 1 - \alpha$ 		\\ \hline
\end{tabular}
\caption{Transformation representations of the superfields for the $U(1)_X = U(1)_{B-L}$ model. 
}  
\label{table:BLuv}
\end{table}

The $B-L$ charges of the baryons $\tB_i$ and $\tbarB_i$ are left in terms of a constant 
$q\neq 0$ which parameterizes their size relative to the canonical charges of the MSSM matter. 
While generic values of $q$ are phenomenologically viable, certain choices would
permit low-dimensional \upq-violating operators and spoil the axion quality.
The problematic $q$ can be identified by considering all of the low-dimensional $SU(5)_\text{SM}$ singlet operators with nonzero $B-L$ charge:
\begin{align}
(\bar\nu_R)_{-1} &,&
(\bar\nu_R^n)_{-n} &,&
(L \tbarM^{(2)})_{+1} &,&
(\bar d_R \tM^{(3)} )_{-1/3} &,&
(\tbarM^{(3)} Q_L L)_{+1/3} ,
\end{align}
where the subscripts indicate the $B-L$ charge of each operator.  Since none of these carry PQ charge,
the superpotential operator constructed by multiplying any of them by a baryon superfield would violate $\upq$ unacceptably.
To avoid this issue, we restrict ourselves to the cases where $q \neq \pm \frac{n}{5}$, for $n=0,1,2,3,4$, and also $q \neq \pm \frac{1}{3}$.

\subsubsection{Composite Higgs Doublets}

The identification of $X = B - L$ has positive implications for the superpotential, notably by forbidding
 many of the operators that would mediate highly constrained 
$B$ and/or $L$ violation such as proton decay~\cite{Font:1989ai}. 
The allowed low energy effective superpotential has the form:
\begin{equation}
W =  \mu \tbarM^{(2)} \tM^{(2)}  + \mu' \tbarM^{(3)} \tM^{(3)} + y_u Q_L \tM^{(2)} \bar u_R  
+ y_d Q_L \tbarM^{(2)} \bar d_R + y_e L \tbarM^{(2)} \bar e_R + y_\nu L \tM^{(2)} \bar \nu_R,
\label{eq:Wmssm}
\end{equation}
containing mass terms for the doublet and triplet mesons, and Yukawa interactions for the doublets with the MSSM matter.

The mesons $\tbarM^{(2)}$ and $\tM^{(2)}$ have the same gauge representations as the MSSM Higgs superfields $H_d$ and $H_u$.
We take the economical route of interpreting the lightest $\tbarM^{(2)} + \tM^{(2)}$ pair of the five flavors of $SU(2)_L$ doublet mesons
as composite MSSM Higgs superfields, which potentially offers insight into the $\mu$ problem of the MSSM.
The terms in \eqref{eq:Wmssm} descend from non-renormalizable composite operators in the UV theory.  In the case of the
$\mu$~terms, these operators are dimension-4 and violate the $U(1)_R$ symmetry.  If generated by quantum gravitational residuals,
the natural mass scale for $\mu$ and $\mu'$ would thus be:
\begin{equation}
W_g \sim \frac{(\barQ_2 \barQ_1) (Q_1 Q_2)}{\mpl} \longrightarrow \frac{\Lambda_L \Lambda_R}{\mpl} \left( \tbarM^{(2)} \tM^{(2)} +  \tbarM^{(3)} \tM^{(3)} \right) \longrightarrow \mu, \mu' \sim \frac{\Lambda_L \Lambda_R}{\mpl}.
\label{eq:Wgmu}
\end{equation} 
This is $\mu \sim \mathcal O(\tev)$ for our benchmark choice of $\Lambda_L \approx \Lambda_R \approx 10^{11}\ \gev$.  

The Yukawa interactions of \eqref{eq:Wmssm} similarly correspond to dimension five operators in the UV.  Realizing the large couplings necessary for the heavy
quarks requires that they be generated at a lower scale $M_F \ll \mpl$:
\begin{equation}
W = y_u' \frac{Q_L (Q_1 Q_2) \bar u_R}{M_F} + y_d' \frac{Q_L (\barQ_2 \barQ_1) \bar d_R}{M_F} + y_e' \frac{L (\barQ_2 \barQ_1) \bar e_R}{M_F} , \label{eq:WyudR}
\end{equation} 
where $y_t \sim 1$ requires $M_F \sim \Lambda_R$ (and $y_b$ requires $\Lambda_L$ is not much larger).
Unlike the dynamics generating the $\mu$ terms, the Yukawa interactions are compatible with the $U(1)_R$ symmetry,
which allows for the disparate scales to remain technically natural.

The presence of the four additional $\tM^{(2)}$ and $\tbarM^{(2)}$ in \eqref{eq:WyudR} poses a potential phenomenological problem. 
In the absence of any additional structure, the $y_{u,d,e}'$ couplings of the matter fields with the heavier $SU(2)_L$ doublets will generally 
introduce flavor-changing neutral currents (FCNC). 
A number of potential solutions exist in the literature. For example, by imposing minimal flavor violation~\cite{Ciuchini:1998xy} 
on \eqref{eq:WyudR}, the $\tM^{(2)}$ and $\tbarM^{(2)}$ can have masses as small as a few $\tev$. Or, as we 
discuss in Section~\ref{sec:breaking}, a discrete symmetry can be imposed (even if broken at $\mpl$)
to forbid the $y_{u,d,e}'$ couplings for all of the mesons except for $H_u$ and $H_d$.

\subsubsection{Color-Triplet Mesons}

As illustrated in \eqref{eq:Wgmu}, we expect that gravitational effects induce electroweak scale
$\mathcal O(\frac{\Lambda_L \Lambda_R}{\mpl})$ supersymmetric masses for each of the five pairs of $\tbarM^{(3)} \tM^{(3)}$ color triplets. 
Generically, color triplets with weak scale masses are very tightly constrained,
especially because the interactions
\begin{align}
W_\text{bad} \sim Q_L \tbarM^{(3)}  L + \bar u_R \tM^{(3)} \bar e_R + \bar d_R \tM^{(3)} +   \tbarM^{(3)} \bar u_R \tbarM^{(3)} + \ldots,
\label{eq:Wbad}
\end{align}
if present, would mediate fast proton decay. Fortunately, every term in \eqref{eq:Wbad} is forbidden upon gauging $U(1)_X = U(1)_{B-L}$. 
Thus, $\tM^{(3)}$ and $\tbarM^{(3)}$ are distinct from the Higgs color triplets which 
typically appear in $SU(5)$ grand unified theories. 
In Section~\ref{sec:breaking} we explore the possibility that they could (along with the extra $SU(2)_L$ doublets)
serve as messengers for gauge-mediated supersymmetry breaking.


\subsection{Alternatives to $B-L$} \label{sec:altX}

In addition to $B-L$, there are a number of other acceptable anomaly-free $U(1)_X$ 
charge assignments for the Standard Model matter.  While none are as attractive as $B-L$,
in this section we sketch three alternatives: a ``5/-3/1" pattern of $U(1)_X$ charges within each generation; 
every matter superfield neutral under $U(1)_X$; 
and a $L_i - L_j$ model.

\subsubsection{5/-3/1 Model}

An alternative charge assignment is shown in Table~\ref{table:Xuv}: 
$Q_L$, $\bar u_R$ and $\bar e_R$ fields have $U(1)_X$ charge $q$; $L$ and $\bar d_R$ have charge $-3 q$; and the $\bar \nu_R$ has charge $5q$
to cancel the $U(1)^3_{X}$. anomaly.
Forbidding all \upq-violating operators of dimension less than 10 requires:
\begin{equation}
q \neq \pm 1, \pm \frac{1}{2}, \pm \frac{1}{3}, \pm \frac{1}{4}, \pm \frac{5}{2}, \pm \frac{5}{3}, \label{eq:Xbadq}
\end{equation}
but otherwise $q$ is a free parameter describing a family of models.
With this charge assignment the undesirable baryon and lepton number violating operators 
$L H_u$, $L L \bar e_R$, $Q L \bar d_R$ and $\bar u_R \bar d_R \bar d_R$ are all forbidden, 
and proton decay occurs via the dimension 5 operator $W\sim \bar u_R \bar u_R \bar d_R \bar e_R/\mpl$.

\begin{table}[t]
\centering
\hspace*{-0.75cm}
\begin{tabular}{| c | c | c c c | c | c | c |} \hline
\Tstrut \Bstrut
   		&$SU(5)_1$	&$SU_3$&$SU_2$&$U(1)_Y$	&$SU(5)_2$	&$U(1)_{X}$	& \upq			\\ \hline \Tstrut
$\tbarM^{(3)}$&	\rep{5}	&\repbar{3}&		& $1/3$	&			&$0$			&	0		\\ 
$\tbarM^{(2)}$&	\rep{5}	&		&\rep{2}	& $-1/2$	&			&$0$			&	0		\\ \hline \Tstrut
$\tM^{(3)}$&			&\rep{3}	&		& $-1/3$	&	\repbar{5}	&	$0$		&	0		\\ 
$\tM^{(2)}$&			&		&\rep{2}	& $1/2$	&	\repbar{5}	&	$0$		&	0		\\ \hline \Tstrut
 $\tB_1$, $\tbarB_2$ &	&		&		&	0	&			&	$5 $		&$\pm 1 + \alpha$ 		\\
$\tbarB_1$, $\tB_2$ &	&		&		&	0	&			&	$-5$		&$\pm 1 - \alpha$ 		\\ \hline
$Q_L$	&			& \rep{3}	&\rep{2}	&$1/6$	&			& $+q$		&	0			\\
$\bar u_R$&			& \repbar{3}&		&$-2/3$	&			& $+q$		&	0			\\
$\bar d_R$&			& \repbar{3}&		&$1/3$	&			& $-3q$		&	0			\\ \hline
$L$		&			& 		&\rep{2}	&$-1/2$	&			&	$-3q$		&	0			\\
$\bar e_R$&			&		&		&$+1$	&			&	$+q$		&	0			\\
$\bar \nu_R$&			&		&		& $0$	&			&	$5q$		&	0		\\ \hline 
$H_u$	&			&		&\rep{2}	& $1/2$	&			&	$-2q$	&	0		\\
$H_d$	&			&		&\rep{2}	& $-1/2$	&			&	$2q$		&	0		\\ \hline
\end{tabular}
\caption{Charges of the matter fundamental superfields and Higgs doublets and composite baryons and mesons
in the ``5/-3/1" $U(1)_X$ model. 
}  
\label{table:Xuv}
\end{table}

Unlike in the $B-L$ model, $U(1)_X$ forbids the mesons $\tM^{(2)}$ and $\tbarM^{(2)}$ from having Yukawa interactions
with MSSM matter unless $q = 0$.  Thus, additional fundamental Higgs doublets 
$H_u + H_d$ with $U(1)_X$ charges $\pm 2 q$ must be added to generate quark and lepton masses,
\begin{equation}
W_H = \mu H_u H_d + y_u Q_L H_u \bar u_R +  y_d Q_L H_d \bar d_R +  y_e L H_d \bar e_R +  y_\nu L H_u \bar \nu_R.
\end{equation}
As in the MSSM with fundamental Higgs doublets, there is no \emph{a priori} reason for $\mu$ to be at the weak scale.

Renormalizable couplings between the mesons $\tM$ and $\tbarM$ and the MSSM fields are mediated exclusively 
by gauge interactions.
Direct couplings in the superpotential are suppressed, beginning with the dimension-7 operators $(\tM \tbarM)H_u H_d$.
Direct couplings which would allow the mesons to decay entirely into the Standard Model depend sensitively
on $q$, with the operators permitting prompt decay also typically violating \upq\ and forbidden by \eqref{eq:Xbadq}.
As consequence, the lightest mesons tend to have long lifetimes, and for some values of $q$ can be absolutely stable
and bounded by the strong constraints on colored or charged cosmological relic particles.


\subsubsection{$\mathbf q=0$: Neutral MSSM}

In the limit $q \rightarrow 0$, the MSSM decouples from $U(1)_X$.  This assignment allows for Yukawa interactions
between the mesons and MSSM matter,  permitting $\tM^{(2)}$ and $\tbarM^{(2)}$ to play the role of the MSSM
Higgs doublets, 
with $\mathcal O(\Lambda_L \Lambda_R/\mpl)$ supersymmetric masses as in \eqref{eq:Wgmu}. 
However, $U(1)_X$ no longer forbids the problematic operators of \eqref{eq:Wbad} or
\begin{equation}
W_\text{bad}' \sim L \tM^{(2)} + L L \bar e_R + Q L \bar d_R + \bar u_R \bar d_R \bar d_R.
\label{eq:Wbad2}
\end{equation}
Among the potentially disastrous consequences of $W_\text{bad}'$ is a short proton lifetime. This problem is averted in the MSSM by imposing a $\ZZ_2$ $R$~parity, which ensures that the superpotential respects the $B-L$ global symmetry.
Upon imposing $R$~parity or some other discrete symmetry on the $q=0$ model, the superpotential comes to resemble that of the $B-L$ axion model in all respects except one: if $q=0$ the right-handed neutrino is a singlet under the gauge symmetries, 
at which point it can be safely omitted. 

\subsubsection{$L_i - L_j$ Models}

The Standard Model also admits anomaly-free $U(1)$ symmetries for which charges not are uniform across all three 
generations.  The combinations of $L_\mu - L_\tau$ and $L_e - L_\tau$ are
among the phenomenologically interesting alternatives. 
Models of this type are typically consistent with a composite $H_u$ and $H_d$, 
but as in the MSSM, an $R$~parity must be imposed on such models to ensure 
that all of the $B$ and $L$ violating operators of \eqref{eq:Wbad2} are forbidden.

\section{Gauge-Mediated Supersymmetry Breaking} 
\label{sec:breaking}

Beyond the usual MSSM superfields, there are relatively few additional light degrees of freedom:
\begin{itemize}
\item The four baryons $\tbarB_{1,2}$ and $\tB_{1,2}$ contain at most two light fields in the 
$\ev{\tB_i}\neq0$, $\ev{\tbarB_i} \neq 0$ vacuum. There is a chiral multiplet containing the composite axion.
\item For $U(1)_X$ gauge coupling $g_X \ll 1$, there is a $U(1)_X$ vector supermultiplet with a mass $m_X \sim g_X f_X$,
where $f_X$ is typically $\sim f_a$.
\item The mesons $\tM$ and $\tbarM$ have $\mathcal O(\Lambda_L \Lambda_R /\mpl)$ vectorlike masses. 
In the $B-L$ model and its variants, the lightest such $SU(2)_L$ doublets are identified as the MSSM $H_u$ and $H_d$
leaving four heavier $\tM^{(2)} + \tbarM^{(2)}$ pairs, and five color triplets $\tM^{(3)} + \tbarM^{(3)}$.
\end{itemize}
In this section we explore how these mesons may be utilized as messengers of supersymmetry breaking.

We parameterize the supersymmetry-breaking in a secluded sector as a set of one or more chiral 
superfields $X_i$ acquiring $F$-term expectation values,
\begin{equation}
\ev{X} = \mathcal X + \theta^2 \mathcal F_X,
\end{equation}
with $\mathcal F_X \neq 0$. 
Introducing superpotential terms of the form $W \sim X \tbarM^{(3,2)} \tM^{(3,2)}$
communicates supersymmetry breaking to the MSSM~\cite{Dine:1994vc,Dine:1995ag}.
In the UV theory this superpotential originates from dimension-5 operators 
$(\barQ_2 \barQ_1) X (Q_1 Q_2)/M_S^2$, reducing in the IR to
\begin{equation}
W_s = \lambda_3^{'ij} \left(\frac{\Lambda_L \Lambda_R}{M_S^2} \right) X \tbarM^{(3)}_i \tM^{(3)}_j +  \lambda_2^{'ij} \left(\frac{\Lambda_L \Lambda_R}{M_S^2} \right) X \tbarM^{(2)}_i \tM^{(2)}_j, 
\label{eq:WsM}
\end{equation} 
where the indices $i,j=1\ldots 5$, for some scale $M_S \gtrsim \sqrt{ \Lambda_L \Lambda_R}$ 
which we take to be small compared to $\mpl$. 
It is convenient to absorb the factors of $\Lambda_L \Lambda_R/M_S^2$ into the definitions of $\lambda_{2,3}$:
\begin{equation}
\lambda_{2,3}^{ij} = \frac{\Lambda_L \Lambda_R}{M_S^2} \lambda_{2,3}^{'ij}.
\end{equation}
As with the Yukawa couplings of \eqref{eq:WyudR}, the superpotential $W_s$ respects a global $U(1)_R$ symmetry under which the mesons $\tM$ and $\tbarM$ are neutral, and $X$ has charge $+2$.

As discussed in Section~\ref{sec:BL}, Yukawa-like couplings between the matter fields and the four heavy $\tM^{(2)} + \tbarM^{(2)}$ may introduce unacceptable flavor-changing neutral currents. A standard solution is to impose a ``messenger parity" on the model, under which the Higgs $H_{u,d}$ are even, and the messengers $\tM^{(2,3)}$ and $\tbarM^{(2,3)}$ are odd. Thus, the direct couplings between messenger $SU(2)_L$ doublets and the matter fields are forbidden, and the problematic flavor-changing neutral currents are avoided.\footnote{The messenger parity is a discrete subgroup of the $SU(5)_1 \times SU(5)_2$ flavor symmetry, and can be derived from the breaking pattern $SU(5)_{1,2} \rightarrow SU(4)_{1,2} \times U(1)$ with $\ZZ_2 \subset \ZZ_4 \subset U(1)$, where $H_{u,d}$ and the corresponding $SU(3)_c$ triplets are invariant under the action of $\ZZ_4$.}
Imposing this $\ZZ_2$ symmetry reduces \eqref{eq:WsM} to:
\begin{equation}
W_s = \lambda_3^{1,1} X \tbarM^{(3)}_1 \tM^{(3)}_1 +  \lambda_2^{1,1} X H_d H_u +\sum_{i=2\ldots 5} \sum_{j=2\ldots 5} \left(
\lambda_3^{ij} X \tbarM^{(3)}_i \tM^{(3)}_j +  \lambda_2^{ij} X \tbarM^{(2)}_i \tM^{(2)}_j \right),
\label{eq:WsM2}
\end{equation} 
where, if the messenger parity is derived from the global symmetries of the quarks $Q_{2}$ and $\barQ_2$, we take the $SU(3)_c$ triplets $\tbarM^{(3)}_1$ and $\tM^{(3)}_1$ to be even under the $\ZZ_2$ messenger parity.

Since the mesons come in complete $SU(5)$ multiplets, gauge
unification at a scale $M_\text{GUT}$ is preserved 
due to the fact that $\tM^{(3)} + \tM^{(2)}$ and $\tbarM^{(3)} + \tbarM^{(2)}$ form complete $SU(5)_\text{SM}$ multiplets. Following~\cite{Giudice:1998bp}, 
the gauge coupling strength $\alpha_\text{GUT}$ at the unification scale $M_\text{GUT}$ is modified by
\begin{equation}
\delta \alpha^{-1}_\text{GUT} = - \frac{N_f}{2\pi} \ln \frac{M_\text{GUT}}{\mathcal X}
\end{equation} 
where $N_f=N_c = 5$. 
Requiring that $SU(5)_\text{SM}$ remains perturbative up to the unification scale imposes a lower bound on $\mathcal X$:
\begin{equation}
 \mathcal X \gtrsim 10^{-13} \times M_\text{GUT} \approx 2\ \tev.
\end{equation}

In addition to \eqref{eq:WsM}, the meson messengers also acquire $U(1)_R$ violating mass terms from the 
Planck scale effects, $\mu_{2,3} \sim \Lambda_L \Lambda_R/\mpl$, leading to a scalar mass matrix:
\begin{align}
\left( \begin{array}{c c} \tM_{(2,3)}^{\dagger}  & \tbarM_{(2,3)} \end{array} \right)
\left( \begin{array}{c c} (\lambda_{2,3} \mathcal X + \mu_{2,3})^\dagger (\lambda_{2,3} \mathcal X + \mu_{2,3})	&	(\lambda_{2,3} \mathcal F_X)^\dagger \\
\lambda_{2,3} \mathcal F_X	&	(\lambda_{2,3} \mathcal X+ \mu_{2,3}) (\lambda_{2,3} \mathcal X+ \mu_{2,3})^\dagger \end{array} \right)
\left( \begin{array}{ c} \tM_{(2,3)} \\ \tbarM_{(2,3)}^{\dagger} \end{array} \right).
\end{align}
Performing $SU(4)_{1,2} \times U(1)_{1,2}$ rotations on the fields $\tbarM^{(2)}$ and $\tM^{(2)}$, 
the matrices $(\lambda_{2} \mathcal X + \mu_{2})$ and $(\lambda_{2} \mathcal F_X)$ can be simultaneously diagonalized and made real:
\begin{align}
M_i = (\lambda_{2} \mathcal X + \mu_{2})_{ii}, &&
F_i = (\lambda_{2} \mathcal F_X)_{ii},
\end{align}
with eigenvalues $M_i^2 \pm F_i$.
This basis also diagonalizes the scalar mass matrix of $\tbarM^{(3)}$ and $\tM^{(3)}$ in the special case $\lambda_2 = \lambda_3$ and $\mu_2 = \mu_3$ (but not in general).
Positivity of the (squared) messenger masses imposes a constraint on the $F$-term VEV of the superfield $X$:
\begin{equation}
\mathcal F_X < \frac{\mu_2^2}{\lambda_{2} } + 2\mu_{2} \mathcal X + \lambda_2 \mathcal X^2
\end{equation}
for each pair of $\lambda_{2}^{ii}$ and $\mu_{2}^{ii}$ in the diagonal basis.
Note that due to the compositeness of the messengers, the couplings $\lambda_{2,3}$ are suppressed by a factor $\Lambda_L \Lambda_R/M_S^2$ which may be much smaller than unity. 

To produce the correct electroweak scale, the $M^2$ and $F$ terms for $H_u$ and $H_d$ must coincide. Taking $\lambda_2^{1,1} \sim \frac{\Lambda_L \Lambda_R}{M_S^2}$ and $\mu_2^{1,1} \sim \frac{\Lambda_L \Lambda_R}{\mpl}$, this condition implies a relationship between the scales $M_S$, $\mathcal X$ and $\mathcal F_X$:
\begin{equation}
\mathcal F_X \sim \Lambda_L \Lambda_R \left( \frac{\mathcal X}{M_S} + \frac{M_S}{\mpl} \right)^2.
\label{eq:constrFx}
\end{equation}
Taking the simplifying case $\sqrt{\Lambda_L \Lambda_R} \sim f_a \sim 10^{11}\, \gev$ and 
$M_S \gtrsim f_a$ in the limit $\mathcal X < 10^5\, \gev$, \eqref{eq:constrFx} reduces to the condition $\sqrt{ \mathcal F_X} \sim \frac{f_a M_S}{\mpl}$.
An investigation of the extensions to the composite axion model 
satisfying this constraint would be an interesting opportunity for future work.

\section{Conclusions and Outlook}

We explore a model with a composite axion in which 
an accidental Peccei--Quinn symmetry naturally emerges as a solution to the strong CP problem. 
Gravitational perturbations to the axion scalar potential are shown to be sufficiently suppressed in the $N_c = 5$ model to 
permit an axion decay constant of $f_a \lesssim 3 \times 10^{11}\, \gev$, 
even under the pessimistic assumptions that supersymmetry breaking 
induces the most dangerous \upq-violating $A$-term potential, 
and that the higher-dimensional operators representing quantum gravitational effects
are parameterized by $\mathcal O(1)$ coupling constants.
In addition to providing a satisfactory solution to the axion quality problem, this composite framework is easily extended to any model of axion-like particles (ALPs) with masses much smaller than the scale of spontaneous symmetry breaking.

The general $SU(N)_L \times SU(N)_R \times U(1)_X$ axion model allows the Standard Model matter fields to 
carry nearly any anomaly-free $U(1)_X$ charge assignment without negatively affecting the axion quality. 
In particular, attractive features emerge when $U(1)_X$ is associated with gauging the Standard Model $B-L$ 
global symmetry. The leading terms in the superpotential are those of the MSSM, with none of the problematic 
$B$ or $L$ violating operators that would otherwise need to be forbidden by invoking a discrete ``matter parity".
Additionally, if the Higgs $H_u$ and $H_d$ are taken to be the lightest of the $SU(2)_L$ charged mesons 
from $SU(5)_L$ and $SU(5)_R$ confinement, the dimension-4 gravitationally-induced operator 
naturally generates an electroweak scale $\mu$~term for $f_a \sim 10^{11}$~GeV.
Other choices of $U(1)_X$ charge assignments share this feature, that the $SU(2)_L$ charged mesons have the same quantum numbers as $H_u$ and $H_d$, and could therefore produce a composite Higgs with a $\tev$ scale $\mu$~term.

The low energy phenomenology largely resembles the MSSM plus a chiral superfield containing the standard QCD axion, axino,
and a saxion.  
More unique are the presence of meson fields in vectorlike color triplet and electroweak doublet representations.  
In theories in which the lightest weak doublet pair are identified as the MSSM Higgs superfields, they will have
$\sim$~TeV masses.
Their detailed phenomenology
depends on the $U(1)_X$ charge assignments and some choices of (perhaps slightly broken) global symmetries, and
their presence indicates that the Large Hadron Collider could potentially uncover clues to higher scale physics.
Alternatively, some of these fields could play the role of messengers, leading to a picture in which
supersymmetry-breaking is mediated by gauge interactions.

Among the many opportunities for future work, some promising directions include 
developing the supersymmetry breaking sector, explaining the pattern of Yukawa 
couplings in the MSSM, or exploring the cosmological implications of the composite model in the early universe.

\section*{Acknowledgements}

The authors are grateful for conversations with Arvind Rajaraman, Nathaniel Craig and Csaba Csaki.
This work is supported in part by NSF Grant No.~PHY-1620638, and was performed in part at 
the Aspen Center for Physics, supported by the NSF grant PHY-1607611.

\bibliography{compaxion2}

\end{document}